\documentclass[pdflatex,sn-mathphys-num]{sn-jnl}

\usepackage{graphicx}%
\usepackage{multirow}%
\usepackage{amsmath,amssymb,amsfonts}%
\usepackage{amsthm}%
\usepackage{mathrsfs}%
\usepackage[title]{appendix}%
\usepackage{xcolor}%
\usepackage{textcomp}%
\usepackage{manyfoot}%
\usepackage{booktabs}%
\usepackage{algorithm}%
\usepackage{algorithmicx}%
\usepackage{algpseudocode}%
\usepackage{listings}%

\theoremstyle{thmstyleone}%
\theoremstyle{thmstyletwo}%

\theoremstyle{thmstylethree}%

\begin{document}

\title[Article Title]{Deciphering Spin-Parity Assignments of Nuclear Levels}


\author*[1,2]{\sur{Christian Iliadis}}\email{iliadis@unc.edu}

\affil*[1]{Department of Physics \& Astronomy, University of North Carolina at Chapel Hill, NC 27599-3255, USA}

\affil[2]{Triangle Universities Nuclear Laboratory (TUNL), Duke University, Durham, North Carolina 27708, USA}


\abstract{Spin-parity assignments of nuclear levels are critical for understanding nuclear structure and reactions. However, inconsistent notation conventions and ambiguous reporting in research papers often lead to confusion and misinterpretations. This paper examines the policies of the Evaluated Nuclear Structure Data File (ENSDF) and the evaluations by Endt and collaborators, highlighting key differences in their approaches to spin-parity notation. Sources of confusion are identified, including ambiguous use of strong and weak arguments and the conflation of new experimental results with prior constraints. Recommendations are provided to improve clarity and consistency in reporting spin-parity assignments, emphasizing the need for explicit notation conventions, clear differentiation of argument strengths, community education, and separate reporting of new findings. These steps aim to enhance the accuracy and utility of nuclear data for both researchers and evaluators. }

\maketitle

\section{Introduction}\label{sec1}
Imagine navigating with a map where the symbols on the legend are misinterpreted -- mistaking highways for rivers or parks for cities. Your journey might lead you in entirely the wrong direction. This note aims to highlight common misinterpretations of spins and parities as presented in the Evaluated Nuclear Structure Data File (ENSDF). It argues that a significant number of nuclear scientists often misunderstand the information provided, leading to misapplications in research and analysis.

Consider the following example. Suppose, the spin-parity of a level is listed in a research paper as:
\begin{equation}
\label{eq:1}
(1/2, 3/2^+)
\end{equation}
How should this statement be interpreted? Take a moment to reflect on its meaning and reconsider your perspective after reading this note. 

The following sections review the policies in ENSDF (Section~\ref{sec:ensdf}) and the evaluations by Endt and collaborators (Section~\ref{sec:endt}), give  typical examples for spin-parity misinterpretations (Section~\ref{sec:confusion}), and conclude with recommendations (Section~\ref{sec:conclusion}).

\section{ENSDF policies: ``strong'' and  ``weak'' arguments}\label{sec:ensdf}
Propositions for assigning spins and parities to nuclear levels used by ENSDF have been published for many years in the Nuclear Data Sheets (NDS). At this point it is useful for the reader to review these propositions\footnote{See: \url{https://www.nndc.bnl.gov/nds/docs/NDSPolicies.pdf}}. In particular, a distinction is made between ``strong'' and ``weak'' arguments.  The propositions are not repeated here, but a few examples will suffice.

The $40$ strong arguments listed in the Nuclear Data Sheets pertain to experiments involving $\gamma$-ray transitions, $\alpha$ and $\beta$ decays, angular correlations, reactions, etc. For example, strong argument \#26 states:
%
%
\begin{quote}
{\it If the angular distribution in a single-nucleon transfer reaction can be fitted with a unique $L$ value, the spin of the final state $J_f$ is related to the spin of the initial state $J_i$ by $\overrightarrow{J_f}$ $=$ $\overrightarrow{J_i}$ $+$ $\overrightarrow{L}$ $+$ $\overrightarrow{1/2}$.}
\end{quote}


A general feature of the strong arguments is that they are practically independent of nuclear model assumptions. In contradistinction, weak argument \#3 states:
\begin{quote}
{\it Whenever $\Delta J$ $\ge$ $2$, an appreciable part of the gamma transition proceeds by the lowest possible multipole order.}
\end{quote}

Clearly, this argument is based on a model assumption, i.e., that lower values of the $\gamma$-ray multipolarity are more likely than higher values. Without providing additional information, application of this argument may yield reasonable choices for a spin-parity assignment, but they do not yield a firm assignment.

The ENSDF policy of reporting spin-parities makes a distinction between strong and weak arguments: values given without and with parentheses indicate assignments based on strong and weak arguments, respectively. According to their notation, e.g., $4^{(+)}$ means that: {\it Strong arguments show the spin is $4$; weak arguments indicate positive parity.}

Accordingly, if Expression~(\ref{eq:1}) were given in ENSDF, the values and parentheses would have the following meaning: {\it Weak arguments indicate values of $1/2^+$, $1/2^-$, or $3/2^+$.} Notice that  this interpretation does not exclude other possible values of $J^\pi$. 

\section{Endt's policies: ``strong'' arguments only}\label{sec:endt}
The nuclear structure evaluations of Endt and collaborators \cite{ENDT19781,Endt1990,Endt1998}, which cover energy levels in $A$ $=$ $21$ $-$ $44$ nuclei, have been used for many years by a large section of the nuclear physics community. 

Endt and collaborators accepted as a basis for $J^\pi$ assignments the set of propositions, published in the Nuclear Data Sheets, on which {\it strong arguments} are based. However, they also state \cite{ENDT19781}:
\begin{quote}
{\it For the light nuclei the ``weak arguments'', also formulated by the Nuclear Data Group, may cause confusion. They have been disregarded entirely; in other words: ``weak arguments are no arguments''.}
\end{quote}

If Expression~(\ref{eq:1}) were listed in the evaluations of Endt and collaborators, it would have the following interpretation: {\it Strong arguments show that the spin-parity is either $1/2^+$, $1/2^-$, or $3/2^+$.}
The parentheses are used in Endt's notation because the $J^\pi$ value is not unambiguous. Notice, that this range of assignments is exclusive, in contradistinction to the ENSDF notation (Section~\ref{sec:ensdf}). In other words, it does not allow for $J^\pi$ values outside the given range.

\section{Sources of confusion}\label{sec:confusion}
An informal poll among senior colleagues revealed that a majority interprets spin-parity notation—whether in original published papers or evaluations—in the manner described by Endt. 

Confusion arises when interpreting spin-parity statements in original research papers, as the authors typically do not specify which notation convention they followed. For example, consider a research paper reporting a spin-parity range for a nuclear level as Expression~(\ref{eq:1}). Without clarification, it is unclear whether this range is intended as an exclusive choice, following the convention of Endt, or as an inclusive possibility, in line with the ENSDF standard. 

This ambiguity in notation can result in differing conclusions when new measurements are performed, particularly when previously tabulated spin-parity information is used to update constraints based on the new data. For instance, consider a level with its spin-parity listed as Expression~(\ref{eq:1}). If a single-nucleon transfer measurement is subsequently conducted on a zero-spin target nucleus, and the angular distribution is uniquely fitted to an $\ell$ $=$ $1$ value, the interpretation of the prior data becomes crucial.
 
According to Endt's approach, which interprets tabulated spin-parity ranges as exclusive choices, the spin-parity limitation based solely on the strong argument from the new measurement would be listed as ``($1/2^-$, $3/2^-$)'', where the parentheses indicate multiple possible values. Together with Expression~(\ref{eq:1}), which relies solely on strong arguments according to Endt's notation, the updated spin-parity assignment would unambiguously be  $J^\pi$ $=$ $1/2^-$. Therefore, the revised result would be reported using the notation ``$1/2^-$".

However, under the ENSDF convention, which treats tabulated spin-parity ranges as inclusive, the spin-parity restriction based only on the strong argument from the new measurement would be listed as ``$1/2^-$, $3/2^-$'' (without parentheses). Since Expression~(\ref{eq:1}) in ENSDF notation is based on weak arguments, these would be superseded by the strong argument from the new measurement. Consequently, the revised result would be reported as ``$1/2^-$, $3/2^-$'' (without parentheses). 

A comparison of several examples is presented in Table~\ref{tab1}, illustrating how the same experimental $J^\pi$ information for a given level leads to different reported values depending on the adopted notation convention.

\section{Recommendations and conclusions}\label{sec:conclusion}
To mitigate confusion and improve the clarity of spin-parity assignments in nuclear structure research, the following recommendations are proposed:
\begin{enumerate}[1.]
\item Explicitly State Notation Conventions: Authors of research papers should explicitly specify the notation convention they are using for spin-parity assignments. For example, they should clarify whether the range of $J^\pi$ values provided is inclusive or exclusive.
    
\item Differentiate Between Strong and Weak Arguments: Publications should clearly indicate whether spin-parity assignments are based on strong arguments (independent of nuclear model assumptions) or weak arguments (reliant on specific models). 
    
\item Report New Information Separately: Original research papers should clearly distinguish new spin-parity constraints derived solely from their measurements from any prior experimental restrictions, systematic arguments, or theoretical predictions (e.g., shell model calculations). This separation ensures that readers can unambiguously identify the new contribution, facilitating accurate updates to nuclear data evaluations.    

\item Educate the Community on Notation Differences: Given the widespread misinterpretation of notation conventions, ensure that younger scientists and students understand the distinctions between ENSDF and Endt conventions.
\end{enumerate}

\backmatter

\bmhead{Acknowledgements}

I am indebted to Kiana Setoodehnia, Filip Kondev, Robert Janssens, and Caleb Marshall for their valuable feedback. This work is supported by the DOE, Office of Science, Office of Nuclear Physics, under grants DE-FG02-97ER41041 (UNC) and DE-FG02-97ER41033 (TUNL).

\begin{table}[h]

\caption{Comparison of $J^\pi$ notations for identical experimental information.}\label{tab1}%
\begin{tabular}{@{}llll@{}}
\toprule
Experimental information  & ENSDF notation\footnotemark[1] & Endt notation\footnotemark[2]  \\
available  & (strong and weak arguments) & (strong arguments only) \\
\midrule
Based on strong arguments, $J^\pi$ $=$ $5/2^+$ is unambiguous   & $5/2^+$           & $5/2^+$    \\
\midrule
Based on strong arguments, either $J$ $=$ $1/2$ (with               & 1/2, 3/2$^+$          & (1/2, 3/2$^+$)    \\
ambiguous parity) or $J^\pi$ $=$ $3/2^+$                            &                       &   \\
\midrule
Based on strong arguments, $\pi$ $=$ $+$; based on weak             & (1/2, 3/2)$^+$        & $\pi$ $=$ $+$  \\
arguments, $J$ $=$ $1/2$ or $3/2$; other $J$ values not excluded    &                       &                \\   
\midrule
Based on strong arguments, $J$ $=$ $1/2$ or $3/2$; based            &   $1/2^{(+)}, 3/2^{(+)}$  &  ($1/2, 3/2$) \\
on weak arguments, $\pi$ $=$ $+$                                    &                       &   \\       
\midrule
Based on weak arguments, $J$ $=$ $1/2$ or $J^\pi$ $=$ $3/2^+$;      & (1/2, 3/2$^+$)        & no entry  \\
other $J^\pi$ values not excluded                                   &                       &    \\
\botrule
\end{tabular}
\footnotetext[1]{See: \url{https://www.nndc.bnl.gov/nds/docs/NDSPolicies.pdf}.}
\footnotetext[2]{See Refs.~\cite{ENDT19781,Endt1990,Endt1998}.}
\end{table}




\noindent

\bibliography{sn-bibliography}

\end{document}